\documentclass[10pt]{article}
\usepackage[]{emulateapj,times}

\def\hh{H$_2$}
\def\hho{H$_2$O}
\def\oo{O$_2$}
\def\sub #1 {_{\mathrm{#1}}}
\def\ee #1 {\times 10^{#1}}          
\def\ut #1 #2 { \, \textrm{#1}^{#2}} 
\def\u #1 { \, \textrm{#1}}          

\slugcomment{Ap. J. Lett., in press}

\lefthead{Wardle}
\righthead{OH in C-type shocks}

\begin{document}

\title{Enhanced OH in C-type Shock Waves in Molecular Clouds}

\author{Mark Wardle}
\affil{Special Research Centre for Theoretical Astrophysics, \\
University of Sydney, NSW 2006, Australia}


\begin{abstract} 
Cosmic-ray and X-ray ionisations in molecular gas produce a weak 
far-ultraviolet flux through the radiative decay of \hh\ molecules that 
have been excited by collisions with energetic electrons (the 
Prasad-Tarafdar mechanism).  I consider the effect of this 
dissociating flux on the oxygen chemistry in C-type shocks.

Typically a few percent of the water molecules produced within the 
shock front are dissociated before the gas has cooled to 50\,K. The 
resulting column density of warm OH rises from $10^{15}$ to 
$10^{16}\ut cm -2 $ as the ionisation rate is increased from 
$10^{-17}\ut s -1 $ (typical of dark clouds) to $10^{-15}\ut s -1 $ 
(adjacent to supernova remnants).  These column densities produce 
substantial emission in the far-infrared rotational transitions of OH, 
and are consistent with the OH/\hho\ ratios inferred from \emph{ISO} 
observations of emission from molecular shocks.  For high ionisation 
rates the column of warm OH is sufficient to explain the OH(1720 MHz) 
masers that occur where molecular clouds are being shocked by 
supernova remnants.

The predicted abundance of OH throughout the shock front will enable 
C-type shocks to be examined with high spectral resolution through 
radio observations of the four hyperfine ground state transitions of 
OH at 18cm and heterodyne measurements of emission in the FIR (e.g.\ 
from \emph{SOFIA})

\end{abstract}



\section{Introduction}

The energetic electrons produced by cosmic-ray and X-ray ionisations 
in molecular clouds collisionally excite the Lyman and Werner bands of 
\hh.  The subsequent radiative de-excitations generate a weak flux of 
far-ultraviolet (FUV) photons capable of dissociating many molecular 
species (Prasad \& Tarafdar 1983).  Models of chemistry in C-type 
shock waves have neglected the internally generated FUV photons, which 
are able to dissociate $\ga 1$\% of each shock-produced molecular 
species before the gas cools to 50\,K. Although this does not 
significantly modify the abundances of the parent species, the 
relatively small abundance of a dissociation product may be of 
interest.  A particularly important example is the dissociation of 
water to form OH. It is well-established that \hho\ is formed 
efficiently from O\textsc{i} in the hotter part of the shock front by 
endothermic reactions once $T\ga 400 \u K $ (Draine, Roberge \& 
Dalgarno 1983; Kaufman \& Neufeld 1996).  The dissociation of a 
percent of the water implies an OH abundance $\sim 10^{-6}$, and 
column densities $\ga 10^{15}\ut cm -2 $.

This column of warm OH is sufficient to explain the high OH/\hho\ 
abundances inferred from \emph{ISO} observations of shocked molecular 
gas associated with the HH54 outflow (Liseau et al.\ 1996), and the 
supernova remnant 3C391 (Reach \& Rho 1998).  Enhanced OH abundances 
associated with shock waves have also been detected at radio 
wavelengths.  OH(1720 MHz) masers are found where supernova remnants 
are running into molecular clouds (Frail, Goss \& Slysh 1994; 
Yusef-Zadeh, Uchida \& Roberts 1995; Frail et al.\ 1996; Green et al.\ 
1997; Koralesky et al.\ 1998; Yusef-Zadeh et al.\ 1999a,b).  
Population inversion occurs by collisions in warm (40-125\,K), 
moderately dense ($\sim \ut 10 5 \ut cm -3 $) molecular gas in the 
absence of a significant FIR radiation field.  In addition, a 
line-of-sight OH column between $\ut 10 16 $ and $\ut 10 17 \ut cm -2 
$ is required for masing to occur (Elitzur 1976; Pavlakis \& Kylafis 
1986; Lockett, Gauthier \& Elitzur 1999).  The conditions for masing 
in this transition are most likely attained in the cooling tail of a 
C-type shock wave driven into the cloud by the overpressure within the 
remnant (Wardle, Yusef-Zadeh \& Geballe 1998, 1999; Lockett et al.\ 
1999) provided that sufficient water can be dissociated in the shocked 
gas before it cools below $\sim 50 \u K $.

In this \emph{Letter} I present models of the oxygen chemistry in 
C-shocks that include dissociations by internally-generated FUV 
photons.  In \S\ref{sec:models} I outline the physical and chemical 
processes that are included in the calculations, and results for 
ionisation rates typical of dark clouds and clouds adjacent to 
supernova remnants are presented in \S\ref{sec:results}.  Warm OH 
column densities in the range $10^{15}$--$10^{16}\ut cm -2 $ are 
produced, consistent with the presence of OH(1720 MHz) masers if the 
shock front harbouring the masers is propagating roughly perpendicular 
to the line of sight -- as is expected because the masing column 
should have a small line-of-sight velocity gradient.  The implications 
of these results are briefly discussed in \S\ref{sec:discussion}.

\section{Shock Models}
\label{sec:models}

The shock structure is assumed to be steady and plane-parallel with 
the magnetic field perpendicular to the shock normal.  The medium is 
comprised of neutral and ionised fluids, with the magnetic field 
frozen into the ionised component and the ionised and neutral components 
coupled by elastic collisions.  The ionisation fraction is low, so 
the inertia and thermal pressure of the ionised component are 
neglected.  

I follow Kaufman \& Neufeld (1996) by writing the magnitude of the 
drag force per unit volume on the neutrals due to the drift of the 
charged species as
\begin{equation}
	F =  \left(\frac{\rho_i}{\rho_{i0}}\right) \frac{\rho v_d^2}{L} \,,
	\label{eq:drag}
\end{equation}
where $\rho$ and $\rho_i$ are the densities of the ion and neutral 
components, $\rho_{i0}$ is the preshock value of $\rho_i$, 
and $v_d$ is the ion-neutral drift speed.  The drag force is in the 
direction of the ion drift through the neutrals, and there is an equal 
and oppposite force per unit volume exerted on the ions by the 
neutrals.  The coupling lengthscale $L$ is determined by the detailed 
composition of the ionised fluid, a mixture of molecular and metal 
ions, electrons and charged grains.  The calculation of the 
upstream ionisation balance by Kaufman \& Neufeld (1996) yields $L 
\approx 10^{17.2} ( \zeta/10^{-17}\ut s -1 ) (n\sub H / 10^4 \ut cm -3 
) \u cm $ for the densities of interest here.

The ionisation rate $\zeta$ is generally assumed to be due to cosmic 
rays, but X-rays effect molecular gas similarly (e.g.\ Maloney, 
Hollenbach \& Tielens 1996) so both ionisation sources are notionally 
lumped together here.  The cosmic-ray and X-ray ionisation rates 
depend on the environment: in dark clouds the standard interstellar 
cosmic-ray flux yields $\zeta \sim \ut 10 -17 \ut s -1 $, whereas 
$\zeta \sim \ut 10 -15 \ut s -1 $ in molecular gas adjacent to 
supernova remnants.  For example, for a remnant characterised by $L_X 
= 10^{36} \u ergs \ut s -1 $ and a radius of 10 pc, $\zeta \sim 3\ee 
-16 \ut s -1 $ (see \S\ref{sec:discussion}).  The cosmic-ray 
ionisation rate may also be of this order if the low-energy cosmic ray 
flux is increased by a factor of 100 over the local interstellar 
value, as seems to be the case for GeV cosmic rays (Esposito et 
al.\ 1996).  I therefore consider ionisation rates in the range 
$10^{-17}$--$10^{-15}\ut s -1 $.

The ambipolar diffusion heating rate per unit volume, neglecting the 
heat capacity and radiative cooling by the ionised fluid, is $F v_d $; 
heating by PdV work, and ionisations are 
also included.  Cooling is assumed to occur by collisional 
dissociation and vibrational transitions of \hh\ (Lepp \& Shull 1982), 
and by rotational transitions of \hh, \hho, and CO (Neufeld \& Kaufman 
1993; Neufeld, Lepp \& Melnick 1995).

The shock models include a simple reaction network sufficient to 
follow the oxygen chemistry: 
\begin{eqnarray}
	\textrm{\hh + \hh} & \rightarrow & \textrm{H + H + \hh}
	\label{eq:H2diss}  \\
	\textrm{O + \hh} & \rightleftharpoons & \textrm{OH + H}
	\label{eq:O+H2}  \\
	\textrm{OH + \hh} & \rightleftharpoons & \textrm{\hho\ + H}   
	\label{eq:OH+H2} \\
	\textrm{OH + OH}  & \rightleftharpoons & \textrm{\hho\ + O}  
	\label{eq:OH+OH} \\
	\textrm{OH + O}   & \rightleftharpoons & \textrm{\oo\ + H}   
	\label{eq:OH+O} \\
	\textrm{\hho\ + FUV} & \rightarrow & \textrm{OH + H}
	\label{eq:H2O+FUV} \\
    \textrm{ OH + FUV} & \rightarrow & \textrm{O + H}
	\label{eq:OH+FUV} \\
	\textrm{\oo\ + FUV} & \rightarrow & \textrm{O + O} \,.
	\label{eq:O2+FUV}
\end{eqnarray}
Rates for collisional dissociation of \hh\ were obtained from Lepp \& 
Shull (1982); and vibrationally cold rate coefficients for 
(\ref{eq:O+H2})--(\ref{eq:OH+O}) were taken from Wagner \& Graff 
(1987) apart from their forward rate for (\ref{eq:OH+OH}) which is 
ill-behaved below 300\,K -- the rate coefficient from the RATE95 
database (Millar, Farqhar \& Willacy 1997) is adopted instead.  
Photodissociation rates for \hho, OH and \oo\ by internally-generated 
FUV photons were obtained from Gredel et al.\ (1989).

I adopt shock parameters consistent with the OH(1720 MHz) observations 
towards W28, W44, and IC443, which imply post shock densities $\sim 
10^5 \ut cm -3 $, and Zeeman field strengths $\sim 0.3 \u mG $ 
(Claussen et al 1997): a shock speed 25$\u km \ut s -1 $, preshock 
density n$\sub H \textsf{=} \ut 10 4 \ut cm -3 $ and preshock magnetic 
field strength 100 $\mathbf{\mu}$G.

\section{Results}
\label{sec:results}

\begin{figure*}
	\centerline{\epsfxsize=7cm \epsfbox{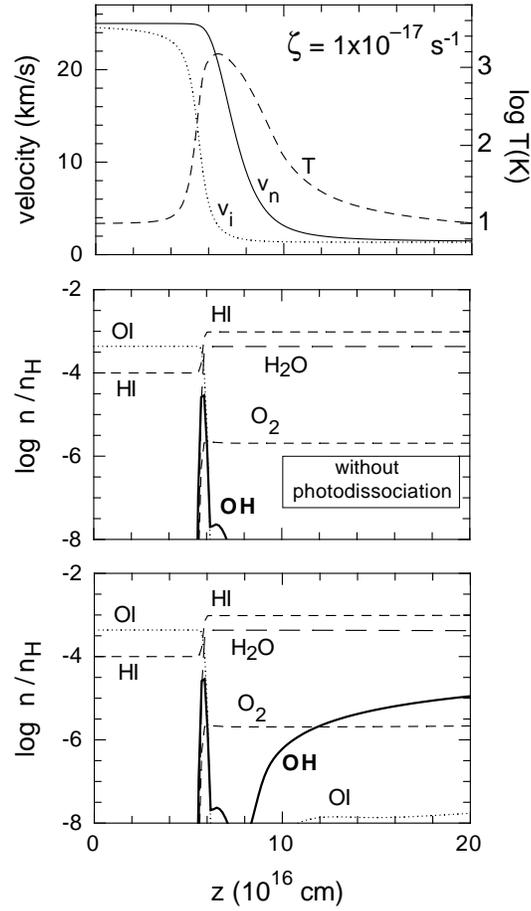}}
	\caption{(\emph{Top panel:} velocity and temperature profile (the 
	temperature scale is on the right) of a $25 \u km \ut s -1 $ 
	C-type shock propagating into gas with $n \sub H = 10^4 \ut cm -3 $ 
	and $B_0 = 100 \mu\u G $.  The X-ray + cosmic ray ionisation rate 
	is assumed to be $\zeta = 10^{-17} \ut s -1 $ (see text).  
	\emph{Middle panel:} oxygen chemistry in the absence of 
	dissociation by the FUV radiation field induced by ionisations.  
	\emph{Lower panel:} oxygen chemistry with dissociations included.
	\label{fig:low_ionisation_rate}}
\end{figure*}

The shock structure for an ionisation rate $\zeta = \ut 10 -17 \ut s 
-1 $, typical of cosmic-ray ionisation in molecular clouds, is 
presented in the top panel of Fig.  \ref{fig:low_ionisation_rate}.  
The velocities are plotted in the shock frame, in which the shock is 
stationary, with the unshocked gas flowing in from $z < 0$ at the 
shock speed (i.e.  $25 \u km \ut s -1 $, being decelerated within the 
shock front and departing downstream towards $z > 0$.  The middle 
panel shows the chemical abundances obtained when dissociations by 
internally-generated FUV photons are neglected.  As has been found 
previously (Draine et al 1983; Kaufman \& Neufeld 1996) the preshock 
atomic oxygen is almost entirely incorporated into water, with OH 
appearing briefly as an intermediate step.  For comparison, the lower 
panel shows the effect of including dissociations by 
internally-generated FUV.  The overall 
chemistry is not drastically affected, but
dissociation of water produces an increasing abundance of OH 
downstream, with a significant column having accumulated by the time 
the gas has cooled to 50\,K. Note also the gradual reappearance of O\textsc{i} 
downstream through the dissociation of OH and \oo.

\begin{figure*}
	\centerline{\epsfxsize=7cm \epsfbox{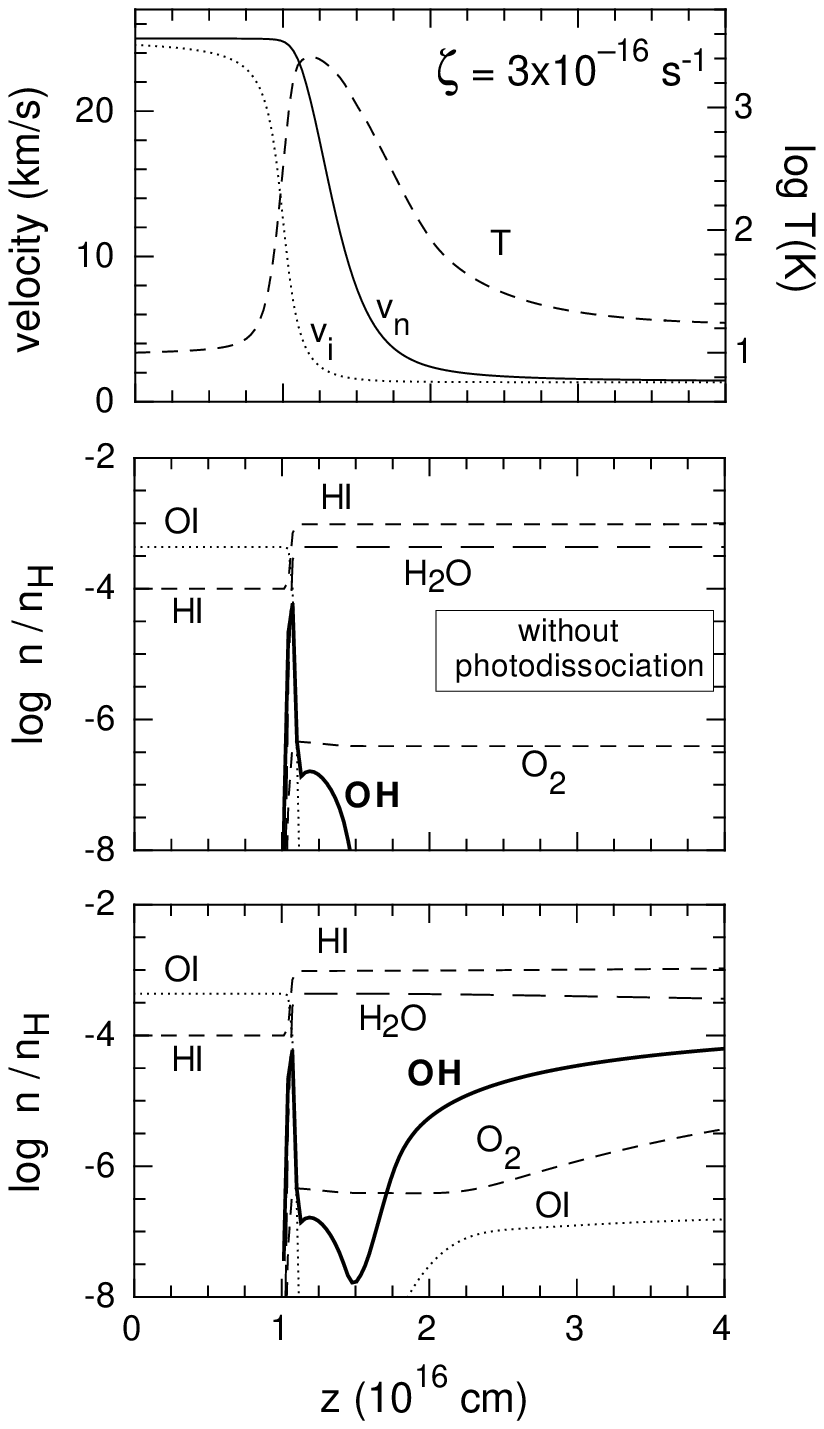}}
	\caption{As for Fig \ref{fig:low_ionisation_rate}, but for an 
	ionisation rate $\zeta = 3\ee -16 \ut s -1 $, typical of molecular 
	clouds  adjacent to supernova remnants.
	\label{fig:high_ionisation_rate}}
\end{figure*}

The effect of raising the ionisation rate to $ 3 
\times \ut 10 -16 \ut s -1 $, as produced by X-rays in a molecular 
cloud adjacent to a SNR, is illustrated in Fig. \ref{fig:high_ionisation_rate}.  
The ionisation fraction of the preshock gas is increased by a factor 
of $\sqrt{30}$, reducing the thickness of the shock transition 
and the time scale for gas to flow through the shock front by the same 
factor.  The peak neutral temperature increases because the energy 
dissipated in the shock must be radiated away on a shorter time scale.  
Heating associated with ionisations, which is 
unimportant within the shock front, increases 30-fold and so the 
final postshock temperature increases slightly.  The photodissociation 
rate is increased 30-fold, with a concomitant increase in the 
production of OH and OI.

\begin{figure*}
	\centerline{\epsfxsize=7cm \epsfbox{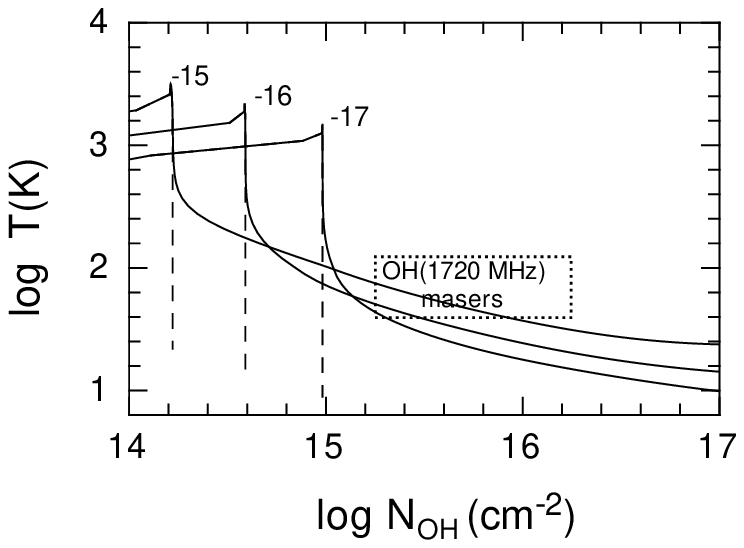}}
	\caption{\emph{Solid:} OH column density versus temperature within 
	C-type shock models with $v_s = 25 \u km \ut s -1 $, $n \sub H = 
	10^4 \ut cm -3 $, and $B_0 = 100 \mu\u G $.  The curves are 
	labelled by $\log \zeta(\ut s -1 )$.  \emph{Dashed:} The same, but 
	with photodissociations neglected.  The dotted box indicates the 
	conditions necessary  for 1720 MHz masers to form assuming that the 
	shock is propagating at $80^{\circ}$ to the line of sight.
	\label{fig:OH_column}}
\end{figure*}

Fig.  \ref{fig:OH_column} summarises the production of OH within the 
shock models by plotting the run of OH column density with temperature 
within the shock front for different ionisation rates.  An initial 
`burst' of OH column is produced as oxygen is rapidly converted to 
water in the high-temperature portion of the shock front.  The column 
produced during this phase decreases as the ionisation rate is 
increased because the OH destroying forward reaction in 
(\ref{eq:OH+H2}) is strongly temperature dependent, and the 
temperature within the shock rises more rapidly for high ionisation 
rates because of the increase in the rate of ambipolar diffusion 
heating.  If the FUV dissociation is neglected no more OH is produced 
(dashed curves); otherwise OH starts to be produced once the 
temperature of the shocked gas has dropped sufficiently so that the 
forward reaction in (\ref{eq:OH+H2}) has slowed sufficiently to permit 
photodissociations to become effective.  For high ionisation rates the 
increase in the photodissociation rate ($\propto \zeta$) is partially 
offset by the decrease in flow time scale ($\propto \zeta^{-1/2}$); 
thus N$\sub OH \propto \zeta^{1/2}$ in the cooling gas.

\section{Discussion}
\label{sec:discussion}

The line-of-sight column density of warm OH through a C-type shock is 
$N\sub OH \sec \theta $ where $\theta$ is the angle between the shock 
normal and the line of sight.  For $\zeta \ga \ut 10 -16 \ut s -1 $ 
and $\theta\ga 80$, N$\sub OH $ is sufficient to permit the 
formation of OH(1720 MHz) masers, thus typical X-ray fluxes near SNRs 
are able to produce the warm OH column necessary for OH(1720 MHz) 
masers.

Lockett et al.\ (1999) have argued that the X-ray flux from SNRs is 
$\la 0.02$ of the level that could produce the OH column density 
necessary for masing in the 1720 MHz line.  This discrepancy can be 
traced to the conversion from X-ray energy flux $F_X$ to ionisation 
rate.  Both Wardle et al.\ (1998) and Lockett et al.\ (1999) used the 
conversion $\zeta( \ut s -1 ) \approx 3\ee -14 F_X(\u erg \ut cm -2 
\ut s -1 )$ (Maloney et al.\ 1996), but this assumes a hard X-ray 
spectrum appropriate to AGNs.  The softer ($\bar{E} \sim 1 \u keV $) 
spectrum from the hot gas in SNRs produces many more ionisations per 
unit X-ray luminosity, as ionisations are dominated by low-energy 
X-ray photons.  In this case, $\zeta \approx N_e \sigma F_X $ where 
$N_e\approx 30 \ut keV -1 $ is the mean number of primary and 
secondary electrons produced by the absorption of unit energy, $\sigma 
\approx 2.6\ee -22 \ut cm -2 $ is the photoabsorption cross-section 
per hydrogen nucleus at 1 keV, and $F_X = L_X/4\pi R^2$.  With $L_X = 
10^{36}\u erg \ut s -1 $ and $R = 10\u pc $, this yields $\zeta 
\approx 4.6\ee -16 \ut s -1 $ provided that the hydrogen column 
density $N_H \la 10^{22} \ut cm -2 $.  Note that externally-incident 
UV may make a similar contribution to the OH column if $A_V$ is not 
too large (Lockett et al.\ 1999).

The enhancement of OH described in this \emph{Letter} permits the 
structure and kinematics of C-type shock waves to be studied through 
the modelling of the excitation of OH throughout the shock front, and 
the calculation of emission and absorption line profiles.  Supernova 
remnants with associated 1720 MHz masers (Frail et al.\ 1994, 1996; 
Yusef-Zadeh et al.\ 1995, 1996, 1999a,b; Green et al.\ 1997; Koralesky et 
al.\ 1998) are obvious observational targets.  At radio wavelengths, 
the four ground-state transitions at 1612, 1665, 1667 and 1720 MHz can 
be observed with sub-km/s velocity resolution in absorption against 
the background continuum from the remnant.  For example, W28 shows 
extended OH absorption around the OH maser positions (Pastchenko \& 
Slysh 1974; Claussen et al.\ 1997).  The predicted warm OH column within 
C-type shocks implies substantial far-infrared emission in low-lying 
OH rotational transitions.  This will, for example, be easily 
detectable by the GREAT heterodyne spectrometer planned for SOFIA, an 
instrument that will be capable of 0.1 km/s or better velocity 
resolution.

\acknowledgements 
I thank Cecilia Ceccarelli, Ewine van Dishoeck, Michael Kaufman and 
Mark Wolfire for discussions, and Phil Lockett for comments on the 
manuscript.  The Special Research Centre for Theoretical Astrophysics 
is funded by the Australian Research Council under its Special 
Research Centres programme.

%
%
%

\end{document}